\documentclass[12pt,preprint,twoside]{aastex}
\usepackage{graphicx}

\shorttitle{LPVs in the LMC from MACHO}
\shortauthors{Fraser et al.}

\begin{document}

\title{The Properties of Long-Period Variables in the Large Magellanic Cloud from MACHO}
\author{Oliver J. Fraser, Suzanne L. Hawley}
\affil{Department of Astronomy, University of Washington \\ Box 351580, Seattle WA, 98195-1580}
\email{fraser@astro.washington.edu, slh@astro.washington.edu}

\author{Kem H. Cook}
\affil{IGPP, Lawrence Livermore National Laboratory \\ MS L-413 \\ P.O. Box 808 \\ Livermore, CA 94550}
\email{kcook@llnl.gov}

\begin{abstract}

We present a new analysis of the long-period variables in the Large Magellanic Cloud (LMC) from the MACHO \emph{Variable Star Catalog}. Three-quarters of our sample of evolved, variable stars have periodic light curves. We characterize the stars in our sample using the multiple periods found in their frequency spectra. Additionally, we use single-epoch Two Micron All Sky Survey measurements to construct the average infrared light curves for different groups of these stars. Comparison with evolutionary models shows that stars on the red giant branch (RGB) or the early asymptotic giant branch (AGB) often show non-periodic variability, but begin to pulsate with periods on the two shortest period-luminosity sequences (3 \& 4) when they brighten to $K_s \approx 13$. The stars on the thermally pulsing AGB are more likely to pulsate with longer periods that lie on the next two P-L sequences (1 \& 2), including the sequence associated with the Miras in the LMC. The Petersen diagram and its variants show that multi-periodic stars on each pair of these sequences (3 \& 4, and 1 \& 2) typically pulsate with periods associated only with that pair. The periods in these multi-periodic stars become longer and stronger as the star evolves. We further constrain the mechanism behind the long secondary periods (LSPs) seen in half of our sample, and find that there is a close match between the luminosity functions of the LSP stars and all of the stars in our sample, and that these star's pulsation amplitudes are relatively wavelength independent. Although this is characteristic of stellar multiplicity, the large number of these variables is problematic for that explanation.

\end{abstract}


\section{Introduction}
\label{intro}
Stellar pulsation of giant stars appears to be a ubiquitous and important phenomenon---RR Lyrae and Cepheid variables form the basis for the distance scales we use. Miras and other long-period variables (LPVs) however, are not as well understood, largely because their cool and tenuous atmospheres are dynamic environments with a great diversity of molecular species forming and disassociating as the star pulsates. In recent years however, these stars have attracted increased attention as micro-lensing surveys of the Large and Small Magellanic Clouds (OGLE --- \cite{1994ApJ...435L.113P}; OGLE II --- \cite*{1997AcA....47..319U}; MACHO --- \cite{macho}) have  produced large catalogs of LPVs. Well-sampled light curves and excellent photometry give us an opportunity to better understand both the mechanisms behind long-period variables and the physical processes at work in the latest stages of stellar evolution.

Before the wealth of data from micro-lensing surveys, LPVs were traditionally classified by the amplitude and stability of their variability in the $V$ band (e.g. The General Catalog of Variable Stars---\cite{GCVS}). In this scheme, stars with well-defined pulsation are classified as Miras if the amplitude of their variability exceeds 2.5 magnitudes in $V$, and as Semi-Regular Type a (SRa) stars if not. Those with multiple periods, unstable periodicity, or poorly expressed periodicity, are classified as SRb stars. The MACHO Project's survey of the Large Magellanic Cloud (LMC) revealed five parallel sequences of LPVs in period-luminosity space \citep{cook96}, prompting a classification scheme that uses the period of pulsation as its primary discriminator. \cite{Wood99} identified the cause of the first three period-luminosity sequences (denoted \emph{A}, \emph{B}, and \emph{C}) as pulsation, and suggested that the two longest period sequences (\emph{E} and \emph{D}) could be attributed to binary systems. The stars in Sequence E showed the characteristic light curves of contact binary systems, and Sequence D stars---those with the longest periods---simultaneously exhibited at least one shorter period that was coincident with Sequence B. This is likely to be the LMC equivalent of the ``long secondary periods'' described by \cite{1963AJ.....68..253H} for Galactic LPVs, although the periods that comprise Sequence D are on average three times shorter than the LSPs listed in \cite{1963AJ.....68..253H}. \cite{Wood99} proposed that these stars are composed of accreting binary systems, with the long period caused by partial eclipses due to an unseen, dust-enshrouded companion.

In this work the LMC period-luminosity sequences will be named in the manner of \cite{fraser05}, from shortest to longest period: 4, 3, 2, 1, E, and D. We retain the names D and E from \cite{Wood99}, but rename his Sequence C to Sequence 1 and count up toward the shorter periods. This approach provides for a graceful way to accommodate additional short-period sequences. Indeed, the use of 2MASS $K_s$ magnitudes as the luminosity indicator caused a split in the original Sequence B---producing Sequences 2 and 3; \cite{2003MNRAS.343L..79K}. A fifth sequence was identified by \cite{2004AcA....54..129S} through the examination of all significant frequencies of these stars instead of just the strongest frequency.

In general, stars brighten and redden as they evolve along the Red Giant Branch (RGB) and the Asymptotic Giant Branch (AGB). Since the typical $J-K_s$ color of the stars in each sequence reddens as we progress from the short-period Sequence 4 to the longer period Sequence 1, this suggests that evolution proceeds from shorter periods toward longer periods, at least for Sequences 1--4 \citep{fraser05}. In fact, the low luminosity bases of Sequences 2, 3, and 4 are heavily populated by RGB stars \citep{2002MNRAS.337L..31I,  2003MNRAS.343L..79K, 2004MNRAS.347L..83K, 2004MNRAS.347..720I}. Above the tip of the RGB, models of AGB stars that include the effects of mass loss confirm that stars continue their evolution to higher luminosities \citep{1993ApJ...413..641V}. In period-luminosity space, the Mira and SRa stars are not clearly separated, with SRa stars found throughout Sequences 1--4.

\cite{2004AcA....54..129S} identified a more useful division for LPVs in the LMC and Small Magellanic Cloud (SMC) than the Mira/SRa/SRb system. In this system an LPV is classified as an OSARG (Ogle Small Amplitude Red Giant) if one of its three strongest periods falls onto Sequence 4, and as an SRV/Mira if not. This division separates LPVs into two groups with a variety of distinct properties, and it also shows that Sequence D is composed of two different populations: a dimmer population that covers a relatively broad period range, and a more luminous, redder population that shows a tighter period-luminosity relationship (the color changes described here can be seen in \cite{fraser05}).

Stars in Sequence E show ``ellipsoidal'' light curves \citep{2004AcA....54..347S}, where the brightness modulation is due to the gravitational distortion of one member of a close binary system. These light curves exhibit dual minima of unequal depths, but the effect is small enough that most methods (including ours) find a period for these systems that is half of the orbital period. We refer to this period as the Fourier period and use it to distinguish Sequences E and D in our plots. When stars in Sequence E are plotted on the period-luminosity diagram at their orbital period (as in \cite{2004AcA....54..347S} and \cite{2006ApJ...650L..55D}) they smoothly join Sequence D. This, along with a recent analysis of OGLE light curves in \cite{2007ApJ...660.1486S}, supports the binary companion explanation for Sequence D put forth by \cite{Wood99}, as does a radial velocity study of several of these stars \citep{2006MmSAI..77..537A}.

In \cite{fraser05} (hereafter Paper I) we used the MACHO and 2MASS magnitudes and colors to characterize LMC stars in each period-luminosity sequence, and classified the stars in each sequence as Miras, SRa, or SRb. At that time, only 52 percent of the stars in the color- and magnitude-defined sample had a well-determined period in the MACHO Variable Star Catalog. In this work, we expand the successfully analyzed stars to 93 percent of our sample, as well as consider their multi-periodic properties. We also use our results to describe the characteristic variability at each of the stages of RGB and AGB evolution by comparison with population synthesis models, including the stars that show very weak or non-existent periodicity.

\section{Data}

The MACHO Project \citep{macho} comprises eight years worth of observations of the Large and Small Magellanic Clouds and the Bulge of the Milky Way. Our sample is drawn from the MACHO LMC Variable Star Catalog \citep{2003yCat.2247....0A}. Sources from the full MACHO database of several million objects were selected for the Variable Star Catalog if the central 80 percent of points in the object's light curve failed to fit a constant magnitude in a $\chi$-squared test.  This criterion resulted in 207,632 candidate variables in the LMC.\footnote{The MACHO Project is available online at: \url{http://wwwmacho.anu.edu.au/}}

\begin{figure}
\begin{center}
\includegraphics[width=0.9\textwidth]{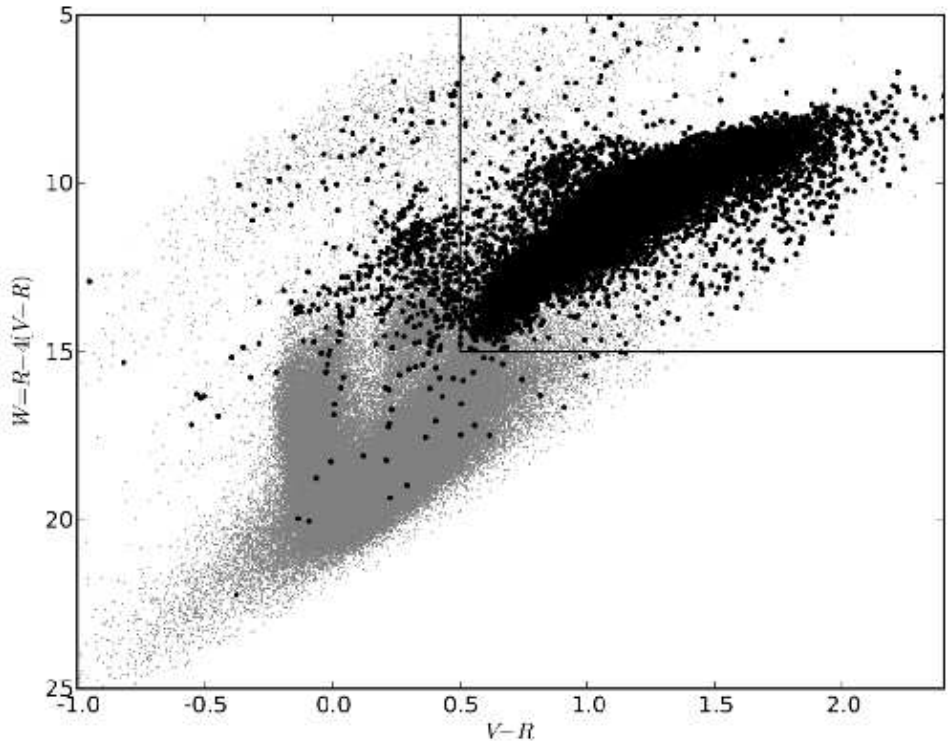}
\caption[The $W$ vs. $V-R$ color--magnitude diagram of the MACHO LMC Variable Star Catalog.]{The $W$ vs. $V-R$ color--magnitude diagram of the MACHO LMC Variable Star Catalog. $W=R-4(V-R)$ is the Wesenheit reddening free magnitude \citep{1995AJ....109.1653A}. The highlighted objects are those identified as LPVs in Paper I \citep{fraser05}, which used the SuperSmoother period and $K_s$ to determine variable type. Our sample is drawn from those stars with $V-R \ge 0.5$ and $W \le 15$. This region surrounds the AGB and encompasses 98 percent of the LPVs from Paper I; the remaining 2 percent, which lie bluer than the present sample, are galactic foreground stars with long SuperSmoother periods. SuperSmoother failed to find a period for approximately half of the stars in this sample, while our technique succeeds 87 percent of the time.}
\label{Sample}
\end{center}
\end{figure}

MACHO data were taken simultaneously in two non-standard filters: Red and Blue. These can be transformed using the method of \cite{1999PASP..111.1539A} to Cousins $V$ and $R$, and then used to find the Wesenheit reddening free magnitude, $W=R-4(V-R)$ \citep{1995AJ....109.1653A}. The construction of $W$ allows very dim stars to enter our sample, so we have removed stars beyond MACHO's dim limits\footnote{Stars with instrumental magnitudes $\ge-1.74$.}. Our LMC sample is defined by $V-R \ge 0.5$ and $W \le 15$ (see Figure \ref{Sample}) and is composed of 56,453 stars. These luminosity and color limits encompass 98 percent of the LPVs from Paper I.

MACHO employed a nonparametric phasing technique known as the SuperSmoother Method \citep{1994PhDT........20R} to phase all of the light curves in the Variable Star Catalog. This technique is robust against complex light curve morphology, but fails when presented with strongly multi-periodic behavior. Multi-periodicity is common among LPVs, and SuperSmoother failed to find a period for nearly half of our sample.

We used the CLEANest algorithm of \cite{1995AJ....109.1889F, 1996AJ....111..541F, 1996AJ....111..555F} to determine the frequency characteristics of the stars in our sample. As implemented by \cite{rorabeck}, and described in \cite{1999ApJ...511..185A}, CLEANest uses the robust date-compensated discrete Fourier transform (DCDFT) algorithm of \cite{1981AJ.....86..619F}, which finds accurate estimates of the amplitudes of the Fourier spectrum for data with uneven time sampling. The CLEANest algorithm iteratively finds the most significant peak in the power spectrum from the DCDFT, adds this frequency to those already known, determines the best fit by allowing all known frequencies to vary slightly, and subtracts the model light curve from the data. The algorithm exits when there is no longer any statistically significant power\footnote{We use a limit of 2, in units of the DCDFT's power, based on the work of \cite{rorabeck} and the suggestion in \cite{1995AJ....109.1889F} that power levels below 2 are ``not even remotely significant.''} in the frequency spectrum. We verified our implementation by checking our results against the 41 MACHO Beat Cepheids from \cite{1995AJ....109.1653A}, and the test dataset from \cite{rorabeck}. CLEANest successfully found every frequency down to our significance threshold.

In the final analysis of our LMC sample, we searched a frequency space of 0.0003 day$^{-1}$ to 0.3 day$^{-1}$ (corresponding to periods from 3.3 days to 3333 days) with a typical frequency resolution of 0.00003 day$^{-1}$. The average number of frequencies returned for a Blue light curve was 13.  Although the MACHO light curves have the necessary time span to uncover frequencies as low as 0.0003 day$^{-1}$, the presence of power at these frequencies is likely caused by slow mean brightness changes. Our analysis found an excess number of frequencies between 0.0003 day$^{-1}$ and 0.0006 day$^{-1}$ (periods between 1666 days and 3333 days). We interpret this as being due to mean brightness changes over the timespan of the MACHO observations, and remove these frequencies from further analysis. 

We employ Two Micron All Sky Survey (2MASS---\cite{2mass}) $K_s$ measurements as our luminosity indicator. The use of infrared luminosities splits Wood's Sequence B into our Sequences 2 and 3. We chose to take all 2MASS matches within two arc-seconds of the MACHO source\footnote{We used an updated astrometric solution for the LMC based on the UCAC system \citep{2000AJ....120.2131Z}.}. This is the distance at which there is a 50 percent chance of a false match. More than 95 percent of our sample has a match within this radius.

Excluding light curves with fewer than 50 points (1 percent of our sample), and stars where the CLEANest analysis failed to converge to reasonable values (8 percent of our sample), the final number of stars with good Red and Blue CLEANest periods is 48,990 (87 percent of our original sample). A comparison of the frequencies found in both the Red and Blue light curves shows that the error in our period estimates is within one-half of a percent up to periods of 55 days, after which it grows to approximately three times our typical frequency resolution (or 0.00009 day$^{-1}$). 

\begin{figure}
\begin{center}
\includegraphics[width=0.9\textwidth]{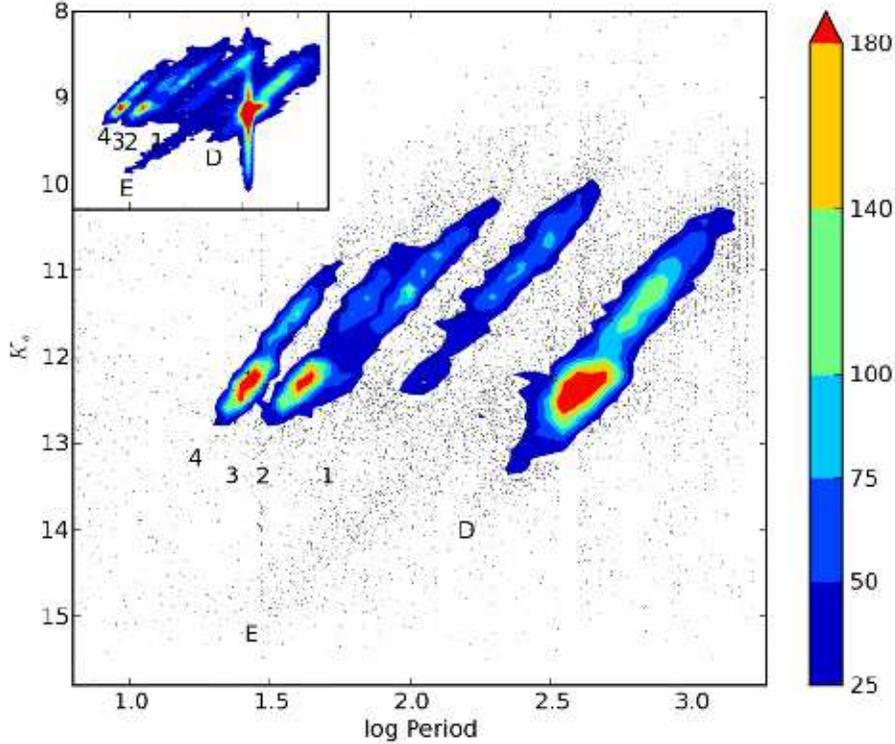}
\caption[Fourier period-luminosity diagram of the LPVs (Long-Period Variables) in the Large Magellanic Cloud]{Fourier period-luminosity diagram of the LPVs (Long-Period Variables) in the Large Magellanic Cloud with and without the stars identified with the One-Year Artifact. Contour levels indicate the density of stars per 0.05 in $\log_{10}\!P_0$ and 0.1 in $K_s$. The Fourier periods shown for stars in Sequence E are half of the orbital periods of these binary systems. When Sequence E is plotted at the orbital period ($+0.30$ in log space) it smoothly joins with the bottom of Sequence D. The inset in this figure shows the Fourier period-luminosity diagram with the One-Year Artifact stars included; the contour levels in the inset start at 5 stars per 0.05 in $\log_{10}\!P$ and 0.1 in $K_s$, and continue at the same contour levels as the larger diagram.}
\label{LMC}
\end{center}
\end{figure}

The primary Fourier period of each star, in days, is the inverse of the first fundamental frequency found in the Blue light curve, $P_0 = 1/\nu_0$. Plotting $K_s$ versus $\log_{10}\!P_0$ (insert of Figure \ref{LMC}) immediately reveals the familiar period-luminosity sequences of the LPVs in the LMC. The sequences are named in the manner of \cite{fraser05}, from shortest to longest period: 4, 3, 2, 1, E, and D. The SuperSmoother and CLEANest analyses of our 56,453 star sample is published in its entirety in the electronic edition of the {\it Astronomical Journal}.

\subsection{Artifact Removal From the Period-Luminosity Diagram}

A very strong vertical feature at periods of one year ($\log_{10}\!P = 2.56$) overlaps both Sequences 1 and D in the Fourier period-luminosity diagram, shown in the inset of Figure \ref{LMC}. This feature is a result of the annual observing schedule of the MACHO Project (\cite{1999PASP..111.1539A}, \S 6.2), and is not an intrinsic property of the star (a similar feature that corresponds to one month is faintly visible in Figure \ref{LMC} at $\log_{10}\!P_0 = 1.49$). Since the strongest period of these stars is not due to the star itself, we associate these 11,215 stars with the One-Year Artifact rather than the sequences with which they overlap in period-luminosity space. Stars associated with the One-Year Artifact are, by their inclusion in the MACHO Variable Star Catalog, variable objects, but they are not necessarily periodic. They certainly have no higher amplitude periodicity in their frequency spectra than the weak signal due to the annual schedule of earthbound telescopes.

In Paper I we simply masked all the stars in this region. In this work we identify just the stars associated with the One-Year Artifact by exploiting differences in the properties of the stars belonging to the artifact with the stars nearby in the Fourier period-luminosity diagram. We identify stars as associated with the One-Year Artifact if they lie in the region $2.53 > \log_{10}\!P_0 > 2.60$, and according to the following rules:

\begin{itemize}
\item Sequence 1 is composed of the reddest stars in our sample. We identify stars as part of the One-Year Artifact if their luminosity places them in or above Sequence 1 and they are bluer than $J-K_s = 1.5$.

\item At luminosities dimmer than Sequence 1 we find that the One-Year Artifact stars have poorly correlated Red and Blue light curves. We identify stars as part of the One-Year Artifact if their Red and Blue periods differ by more than 20 percent, or if the amplitudes corresponding to those periods differ by more than 90 percent. 

\item In the luminosity range between Sequences 1 and D we find that we require the additional parameter of the $\chi^2$ statistic for a sine wave corresponding to $P_0$, which is in the range $2.53 > \log_{10}\!P_0 > 2.60$. We select stars with either Blue amplitudes of less than 0.05 magnitudes mean-to-peak, or those with $\chi^2 < 7 \times 10^4$.
\end{itemize}

Unfortunately the stars in the One-Year Artifact that are dimmer than Sequence D are difficult to differentiate from the background, although certainly the great majority of stars in this region should be identified with the One-Year Artifact. We chose to simply identify stars as members of the One-Year Artifact if they lie below Sequence D and are in the normal period range of $2.53 > \log_{10}\!P_0 > 2.60$.

The One-Year Artifact consists of 11,215 stars with very weak or nonexistent periodicity, or 24 percent of all the stars with good Red and Blue CLEANest results and a 2MASS $K_s$ magnitude. This greatly outnumbers other stars in this narrow period range. Although our process likely does not isolate every star that is associated with the One-Year Artifact, it does serve to uncover what is present below this distracting feature. The Fourier period-luminosity diagram of the LMC with the One-Year Artifact removed is presented in the main panel of Figure \ref{LMC}.

\subsection{ Finding the Average Infrared Light Curves }
\label{irfitting}

\begin{figure}
\begin{center}
\includegraphics[width=0.9\textwidth]{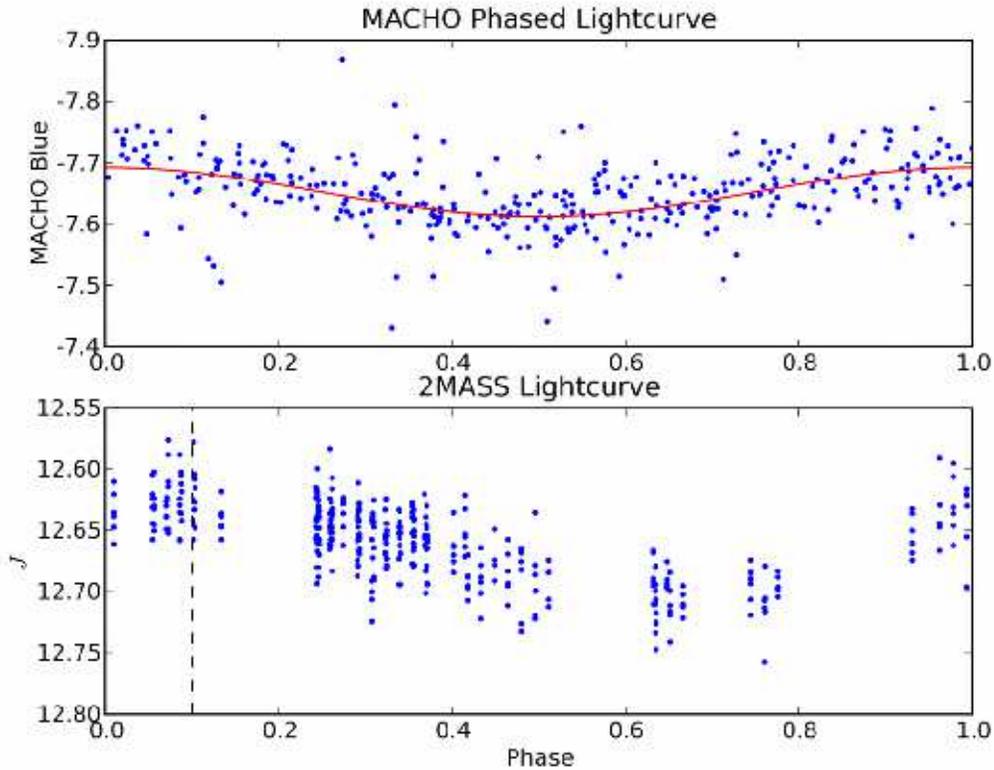}
\caption[Example optical and infrared light curves.]{Example optical and infrared light curves for the star with MACHO designation 55.3126.13. The infrared light curve is taken from the 2MASS calibration tile 90400, and both are phased to the CLEANest period. The infrared light curve shows a phase lag of approximately 10 percent as compared to the optical (the vertical dashed line indicates the approximate maximum of the infrared light curve). This star has an optical period of 64.03 days and lies in Sequence 3. It's optical light curve is shown with the CLEANest model of the primary period; all other frequencies have been subtracted from the model. The average infrared light curve for Sequence 3 was too low in amplitude to measure using our technique, so this star must have unusually strong variation.}
\label{IRLightCurve}
\end{center}
\end{figure}

Although the amplitude of LPVs in the infrared is much lower than in the visible\footnote{In classical pulsators this is due to the black-body behavior of the atmosphere's continuum emission. The large amplitudes of Mira light curves cannot be modeled by this behavior alone; it is also necessary to consider very strong effects from the formation of molecules in the photosphere of the star. See, for example, \cite{2002ApJ...568..931R}.}, there is still intrinsic scatter in the 2MASS observations due to the measurement of each star at a random point in its light curve. Collections of stars with similar light curves can be corrected for this effect in the manner that \cite{2004ApJ...601..260N} used for Cepheid variables. The Fourier period-luminosity sequence in some band is fit with a relation that includes a \emph{correction term} which is a function of $\phi$, the phase of the 2MASS observation with respect to the MACHO light curve. So for each star, $i$:

\begin{equation}
m_i = \alpha \log_{10}\!P_i + \beta + \Omega(\phi_i)
\end{equation}

The correction term, $\Omega(\phi_i)$, is the average infrared light curve for these stars. We chose a form for the correction function based on 2MASS light curves available for 46 stars from our sample. Light curves exist for stars in the 2MASS ``calibration tiles'', fields which were observed multiple times each night to provide photometric calibration. Several tiles lie in the vicinity of the Magellanic Clouds,\footnote{These tiles, 90298, 90299 in the SMC and 90400, 90401, and 90402 in the LMC, were defined to support the 2MASS deep observation campaign of the Magellanic Clouds. Of these five tiles, and the calibration tiles from the rest of the survey, only 90400 and 90401 overlap with MACHO observations.  We only used data from tile 90400 because it was visited more often (377 times, versus 156) over more nights (approximately 40, versus approximately 13) and over a longer time span (90 nights, versus approximately 30) than tile 90401.} but only tile 90400 provided light curves of sufficient length to investigate the behavior typical of long-period variables. An example 2MASS light curve phased to the period of the corresponding MACHO Blue light curve is shown in Figure \ref{IRLightCurve}. For the 46 stars from our sample that match 2MASS sources from this tile, we found that a second-order Fourier series was an adequate model of the light curves in $J$, $H$, and $K_s$.

\begin{equation}
\Omega(\phi) = \sum_{j=1}^2 \ A_j \cos(2 \pi j \phi) + B_j \sin(2 \pi j \phi)
\end{equation}

The phase of the 2MASS observation, $\phi_i$, for each star $i$, is the fractional part of the difference between the time of maximum light, $T_{\mathrm{max},i}$, for the primary Fourier period and $T_{\mathrm{2mass},i}$, the time of the 2MASS observation, in units of the period.

\begin{equation}
\phi_i = \mathrm{mod} \Bigg( \frac{T_{\mathrm{2mass},i} - T_{\mathrm{max},i}}{P_i} \Bigg)
\end{equation}

The model of the infrared light curve, as based on the 46 2MASS light curves, is fit simultaneously with each period-luminosity relationship in our full sample. The scatter about each of these relationships is not completely accounted for, e.g. \cite{2004MNRAS.347..720I} found different relations for  RGB and AGB stars. For this reason we have fit each sequence separately above and below the tip of the Red Giant Branch ($K_s=12.3$ \citep{2000ApJ...542..804N}) for Sequences 2, 3, and 4. Additionally, the widely varying properties of the stars that are being combined tends to reduce the amplitude of the average light curve. We take the limit of detectability of the infrared light curves to be a mean-to-peak amplitude of 0.02 magnitudes---the typical magnitude error of our 2MASS observations---and further require that the scatter about the period-luminosity relationship narrows with the addition of the infrared light curve. At this level we detect infrared light curves in $J$, $H$, and $K_s$ for Sequence 2 above the tip of the RGB, and in Sequences 1, E and D. The fitting and correction was performed individually for two groups in Sequence D: those with Blue $P_0$ peak-to-peak amplitudes less than or equal to 0.2, and those with amplitudes greater than 0.2. This separation roughly corresponds to the division of LMC stars in OGLE by \cite{2004AcA....54..129S} into OSARGs and SRV/Miras. In Sequence 1 we found it necessary to create three amplitude bins (with boundaries at 0.4 and 1) as well as separating the oxygen and carbon stars (using the color cut $J-K_s=1.4$). This created a total of six groups in Sequence 1 that were each fit individually. The results of these fits are tabulated in Table \ref{irfits}, and the infrared light curves are discussed in \S \ref{irlightcurves}.

In \cite{2004ApJ...601..260N} the remaining scatter around the period-luminosity relationship for LMC Cepheids was used to derive individual distances and extinctions for each star. In that work, performing the original fit again after fitting for the distance and extinction together yielded a closer fit in all bands. We have tried this technique in $J$, $H$, and $K_s$ using the scatter around the Fourier period-luminosity relationship in $W$ as a test of its success ($W$ cannot be used to constrain extinction since it is constructed to remove the effects of reddening). Unfortunately the dispersion in $W$ failed to improve after this step, leading us to believe that the extinction in these stars requires a more careful model than simply relying upon the color excess. In a separate test, we found that the addition of a color term also failed to improve the dispersion in the unrelated Fourier period-luminosity relations. LPVs must have some intrinsic luminosity variation that is not well modeled simply by using pulsation with color or individual distances and extinctions. 

Our infrared light curves have very low amplitudes due to the varying properties of their constituent stars. Because the correction is so small, any individual star's departure from the average light curve is likely to dominate the effect of the random phase of the 2MASS observation, so we have not applied these corrections to the 2MASS magnitudes used in this work. While the improvement of the Fourier period-luminosity fits is modest, the infrared light curves do prove useful in providing physical insight into the variability mechanisms as described in \S \ref{irlightcurves} below.

\section{Results}
\label{results}

The Fourier period-luminosity diagram of the LMC with the One-Year Artifact removed is presented in Figure \ref{LMC}. The sparsely populated fifth sequence found by \cite{2004AcA....54..129S}, which lies toward shorter periods of Sequence 4, is not seen in this diagram. This sequence is only seen in the secondary periods ($P_1$), and we have plotted just the primary Fourier period ($P_0$) for each star. All of the stars that compose the fifth sequence have stronger periods elsewhere on the Fourier period-luminosity diagram, primarily on Sequence 4 or the One-Year Artifact. 

Stars in this diagram were grouped into sequences based on the contour traced at a density of five stars per 0.05 in $\log_{10}\!P_0$ and 0.1 in $K_s$ (as shown in the inset of Figure \ref{LMC}). The stars in the high-luminosity tip of Sequence 1 are carbon stars \citep{2004A&A...425..595G}, some of which are heavily self-extincted and fall into the gap between Sequences 1 and D. Such stars are easy to recognize due to their heavy reddening \citep{2000ApJ...542..804N}, thus stars in the gap between Sequences 1 and D with $J-K_s>1.4$ are assigned to Sequence 1. 

Sequence E is well known to merge with Sequence D when plotted at its true orbital period \citep{2004AcA....54..347S, 2006ApJ...650L..55D}, but we have chosen to plot Sequence E at its Fourier period like the other stars on this diagram. This allows us to differentiate these stars from those at the bottom of Sequence D, but it presents us with the problem of separating the two sequences. It's clear that Sequence E in Figure \ref{LMC} is poorly populated, and at the top of the sequence ($K_s < 13$) it's difficult to tell where Sequence E exists among the background of stars between Sequences 1 and D. We chose limits between these sequences such that Sequence E would remain sparsely populated at its high-luminosity end.

\begin{figure}
\begin{center}
\includegraphics[width=0.9\textwidth]{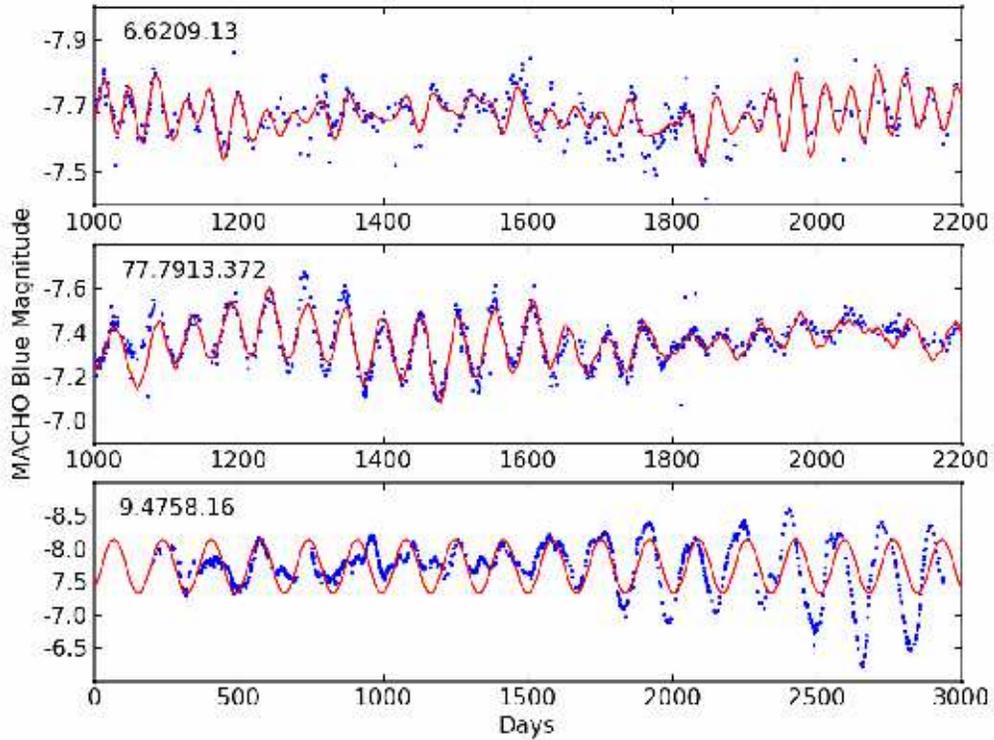}
\caption[Example light curves and CLEANest models of LPVs from our sample.]{Example light curves and CLEANest models of LPVs from our sample; each panel is labeled with the star's MACHO designation. The top panel shows a star from Sequence 4 with close Fourier periods and a sufficiently complicated light curve that a period close to 1 day was falsely assigned in Paper I. The second panel shows another star from Sequence 4 that exhibits amplitude modulation due to its beating periods of 52.18 and 53.40 days. The bottom panel shows the entire MACHO light curve of a Mira variable from Sequence 1. This star's periodicity changes character over the course of the survey, and it has a CLEANest spectrum with beating periods of 168.55 and 169.58 days. A representative sine curve at the primary CLEANest period is shown.}
\label{LightCurves}
\end{center}
\end{figure}

\begin{figure}
\begin{center}
\includegraphics[width=0.9\textwidth]{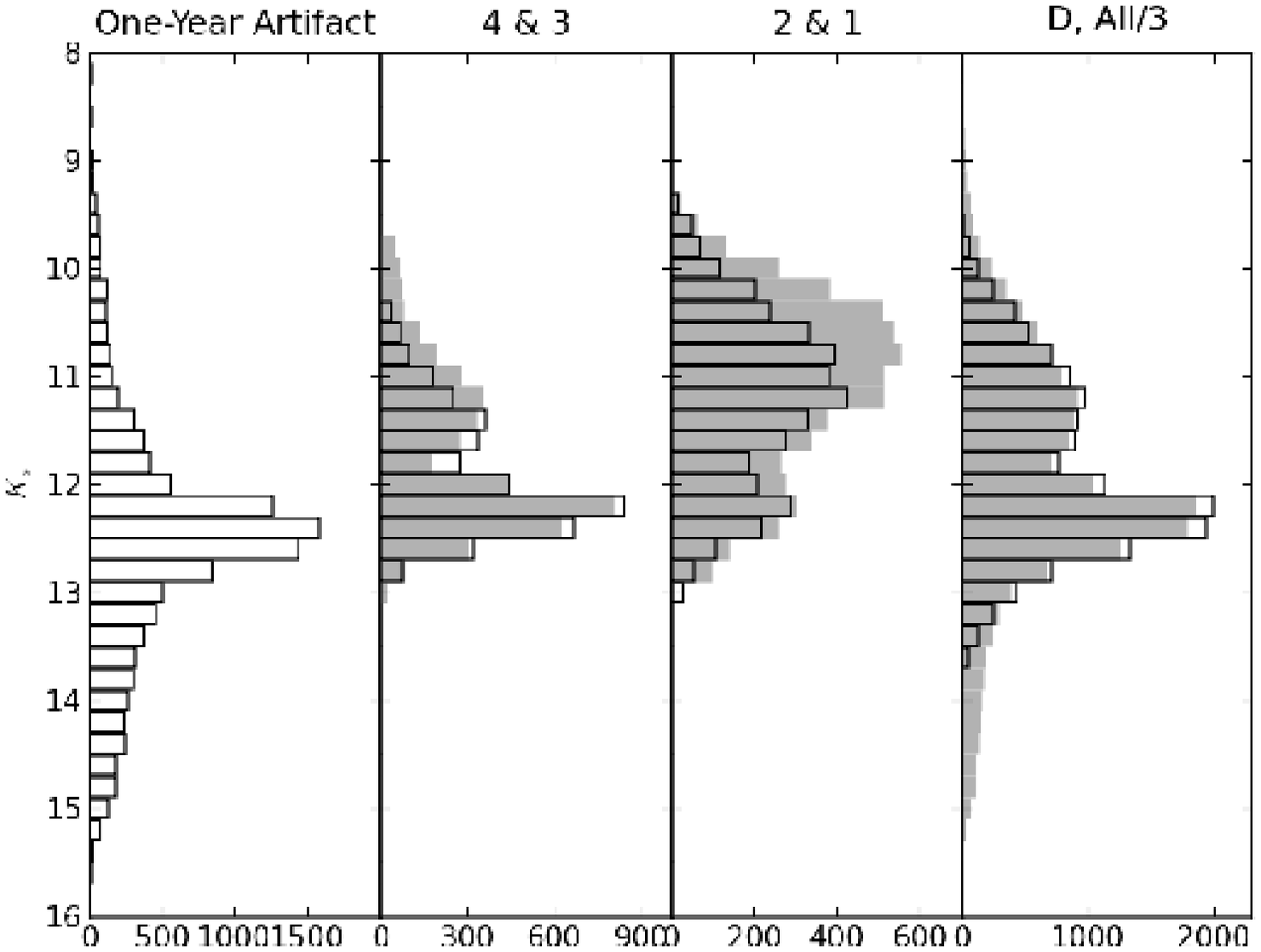}
\caption[Luminosity functions of selected groups of stars from our sample.]{Luminosity functions of selected groups of stars from our sample. The panel on the far left shows the luminosity function of the One-Year Artifact. The second panel compares Sequence 4 (black outlined bars) with Sequence 3 (gray bars with no outlines). The third panel similarly compares Sequences 2 (black outlined bars) and Sequence 1 (gray bars with no outlines). Finally, the panel on the far right compares the luminosity functions of Sequence D (black outlined bars) with our sample (gray bars with no outlines, scaled by $1/3$).}
\label{LumFunctions}
\end{center}
\end{figure}

% Sequence  paperI%   now
%        4        4     8  <-
%        3        7     9  <-
%        2        9     8
%        1       12    11
%        E        4     2
%        D        9    32  <--
%  aliases       36    22  <-
%  `type-0'
Compared to Paper I, where half of that sample had no period assigned by the SuperSmoother analysis, or a period assigned to one year or multiples of one day, we have added approximately 20,000 stars to the period-luminosity diagram. The newly analyzed stars lie mostly on the sequences with the lowest amplitudes (D, 4, and to a lesser extent, 3) and are below the tip of the RGB ($K_s = 12.3$, \cite{2000ApJ...542..804N}). An example is shown in Figure \ref{LightCurves}, top panel. Sequence E stars were well represented in Paper I even though their light curves also have low amplitudes. Half as many stars are identified with the One-Year Artifact due to the careful identification of these stars, as opposed to masking all stars with in this region. Luminosity functions for many of the sequences are shown in Figure \ref{LumFunctions}.

Sequence D as presented in Paper I was underrepresented, accounting for only 9 percent of LPVs. It now represents 31 percent of the stars with good Red and Blue CLEANest results and a 2MASS $K_s$ magnitude (Table \ref{sequences}). The right panel of Figure \ref{LumFunctions} shows the close match between the luminosity functions of Sequence D and our sample, suggesting that stars on Sequence D are drawn from the entire population of long-period variables in the LMC.

The amplitudes of the light curves that are quoted in this work are the peak-to-peak amplitudes found by CLEANest for the $P_0$ term. CLEANest amplitudes underestimate the actual light variation since some of the power associated with this period is contained in harmonics and mixing terms. The amount by which the amplitudes are underestimated is found by comparing the CLEANest amplitudes to amplitudes listed in the MACHO Variable Star Catalog for stars with SuperSmoother periods. SuperSmoother amplitudes were calculated by finding the difference between the mean magnitudes of the closest points to the maximum and minimum of the SuperSmoother phased light curve. On average, the ratio of the SuperSmoother amplitude to the CLEANest amplitude is $1.5$ for stars in Sequence E and stars in the long-period edge of Sequence 1, while the remaining stars on the Fourier period-luminosity diagram tend to show ratios of $1.7$. We have not applied a correction factor to our CLEANest amplitudes, and use them only for relative comparisons.

\begin{figure}
\begin{center}
\includegraphics[width=0.8\textwidth]{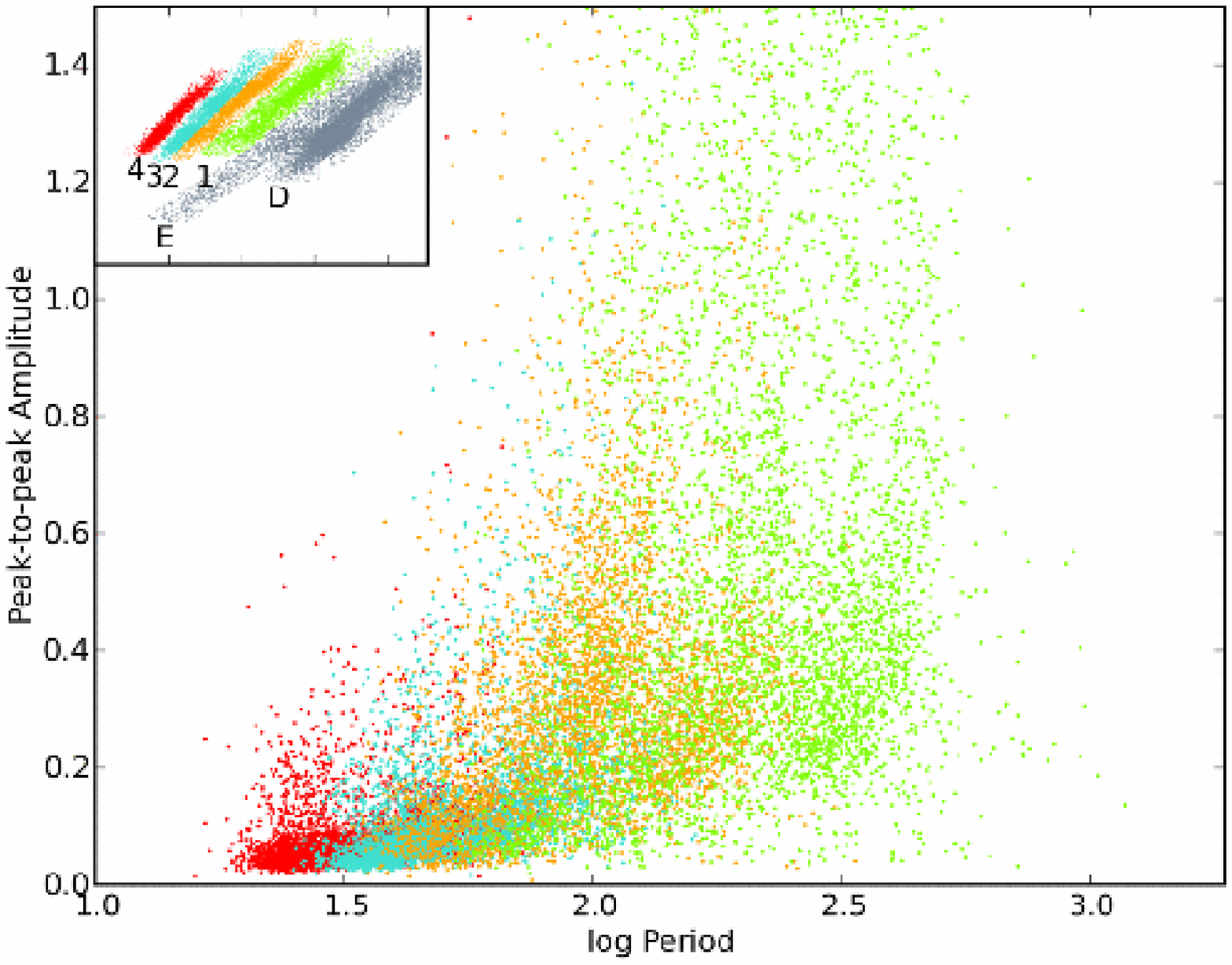}
\includegraphics[width=0.8\textwidth]{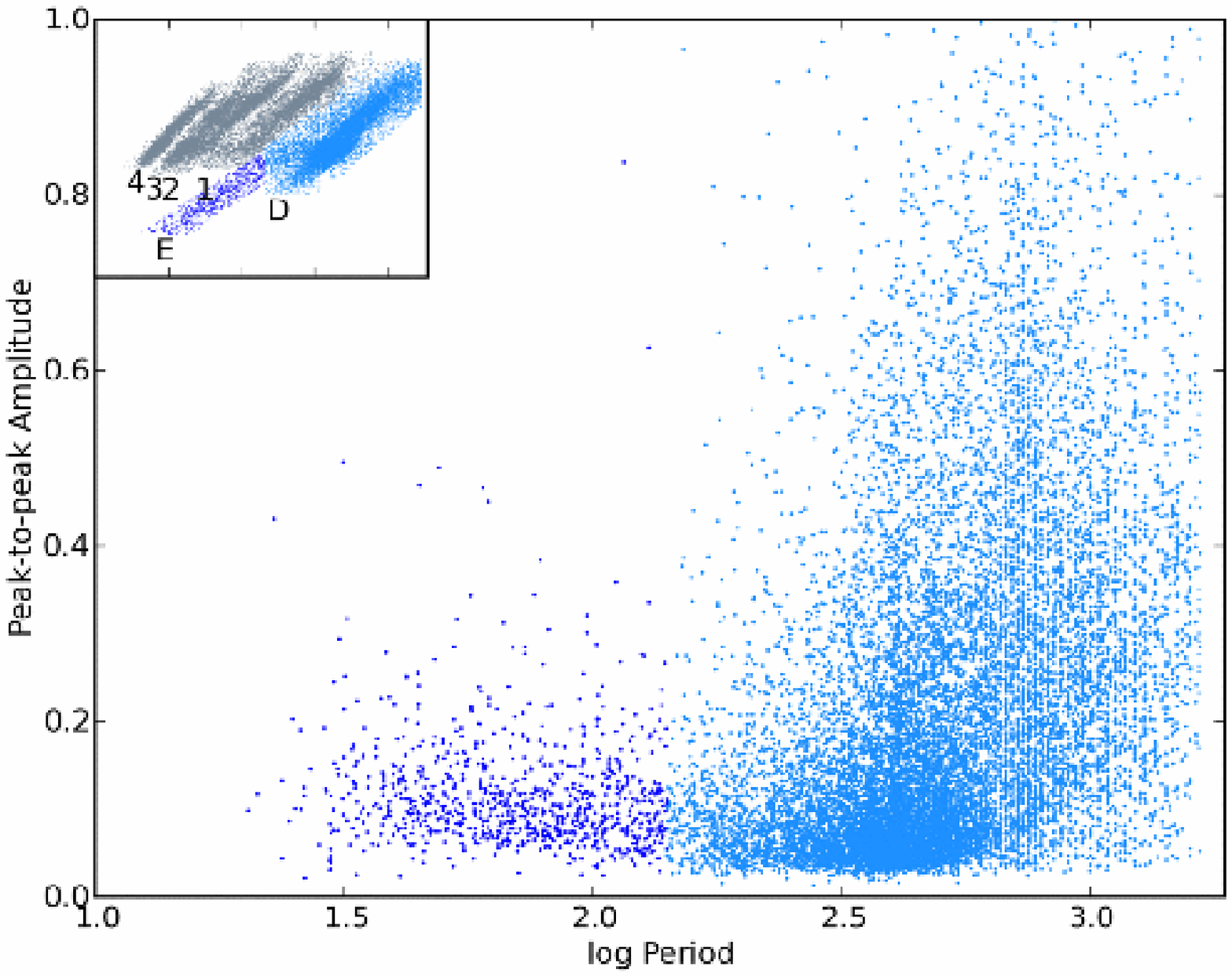}
\caption[Peak-to-peak MACHO Blue amplitude versus $\log_{10}\!P_0$.]{Peak-to-peak MACHO Blue amplitude versus $\log_{10}\!P_0$ for stars on Sequences 1--4, and E and D. Stars are color-coded according to the sequence on which their primary Fourier period lies, as shown in the inset. The Fourier periods shown for stars in Sequence E are half of the orbital periods of these binary systems, a difference of $+0.30$ in log space.}
\label{chapter3Amps}
\end{center}
\end{figure}

The relationship between the pulsation amplitude and $\log_{10}\!P_0$ is markedly different for the different sequences. Figure \ref{chapter3Amps} shows that for stars in Sequences 1--4 there is a correlation between increasing period, increasing amplitude, and increasing range in amplitudes. As compared to stars on Sequences 3 and 4, stars on Sequences 1 and 2 pulsate with much higher amplitudes. Sequence E, as expected for binary stars, does not show a strong dependence of amplitude on period. This sequence is a clear continuation of the lower amplitude group of Sequence D stars, those related to the OSARGs. The moderate amplitude-luminosity correlation for stars in Sequence D reported by \cite{2006ApJ...650L..55D} is not seen for the bulk of the small-amplitude population of Sequence D, using $\log_{10}\!P_0$ as a proxy for luminosity.

\subsection{The Average Infrared Light Curves of LPVs}
\label{irlightcurves}
% NOTE that the visual amplitudes are mean-to-peak
% double them in publication.
% Sequence  1/2Visual Amp  Ks Amp    Ratio
%        4         0.04     0.03       -
%        3         0.07     0.02       -
%        2rgb      0.09     0.02       -
%        2agb      0.21     0.04      11
%        1O low    0.11     0.06       4
%        1O mid    0.32     0.07       9
%        1O high   0.98     0.26       8
%        1C low    0.14     0.08       4
%        1C mid    0.31     0.08       8
%        1C high   0.90     0.36       5
%        E         0.06     0.05       2
%        D down    0.05     0.05       2
%        D up      0.24     0.06       8
%  aliases         0.07        -       -
%  `type-0'        0.17

The comparison of the average infrared light curves with the corresponding optical light curve's properties is a useful constraint on variability mechanisms in these stars. In many types of pulsating stars---including Miras---pulsation has a stronger effect in the optical than the infrared, while variation due to binary systems show light curves of similar amplitudes in all wavebands. Although the average amplitudes of infrared light curves are smaller than the specific star's amplitudes that they are fit to, comparisons can be made in a relative sense. While Table \ref{irfits} presents information for all of the infrared light curves, a summary of amplitude ratios and phase lags for the average Blue and $K_s$ light curves is presented in Table \ref{sequences} for each of the sequences.

We find a clear division in the Blue/$K_s$ amplitude ratios between Sequences 1 and 2, and Sequence E. As expected for binary systems, Sequence E has a smaller amplitude ratio and it has similar amplitudes among the $J$, $H$, and $K_s$ light curves. The pulsating stars in Sequences 1 and 2 show higher ratios and decreasing amplitude with redder wavebands. Interestingly, only the low-amplitude group of Sequence D stars is similar to Sequence E. Although the higher amplitude group in Sequence D has similar amplitudes among the 2MASS bands, the average Blue amplitude is much higher, leading to an amplitude ratio of eight, which is most similar to Sequence 1.

Pulsation modes in LPVs can also be constrained using the phase lag between the optical and the infrared light curves. \cite{2006AJ....131..612S} searched for such phase lags using data from the DIRBE instrument on the COBE satellite. In their sample of 21 stars, all of the Miras---including one carbon star---showed phase lags, while four of the five SR variables did not. They compared the stars in their sample to time-resolved dynamical models of oxygen-rich stars from the literature, and found that phase lags are predicted for fundamental-mode oxygen rich stars, but not for stars pulsating in the first overtone mode. The carbon star models available at the time did not consistently predict a phase lag. Our average infrared light curves show phase lags of 10--20 percent for both Miras and SR variables. The example light curve from the 2MASS calibration tiles, shown in Figure \ref{IRLightCurve}, demonstrates this phase lag. The average infrared light curves for the $0.4 < \mathrm{amplitude} < 1$ bins in Sequence 1 are the exceptions to this rule. Neither the oxygen stars nor the carbon stars in this range show the typical phase lag commonly observed among other LPVs. However, the most notable exception is Sequence D, both its low-amplitude and high-amplitude infrared light curves \emph{lead} the optical light curves, unlike any of the other sequences.

\subsection{Multi-Periodic Stars}
\label{multiP}

\begin{figure}
\begin{center}
\includegraphics[width=0.9\textwidth]{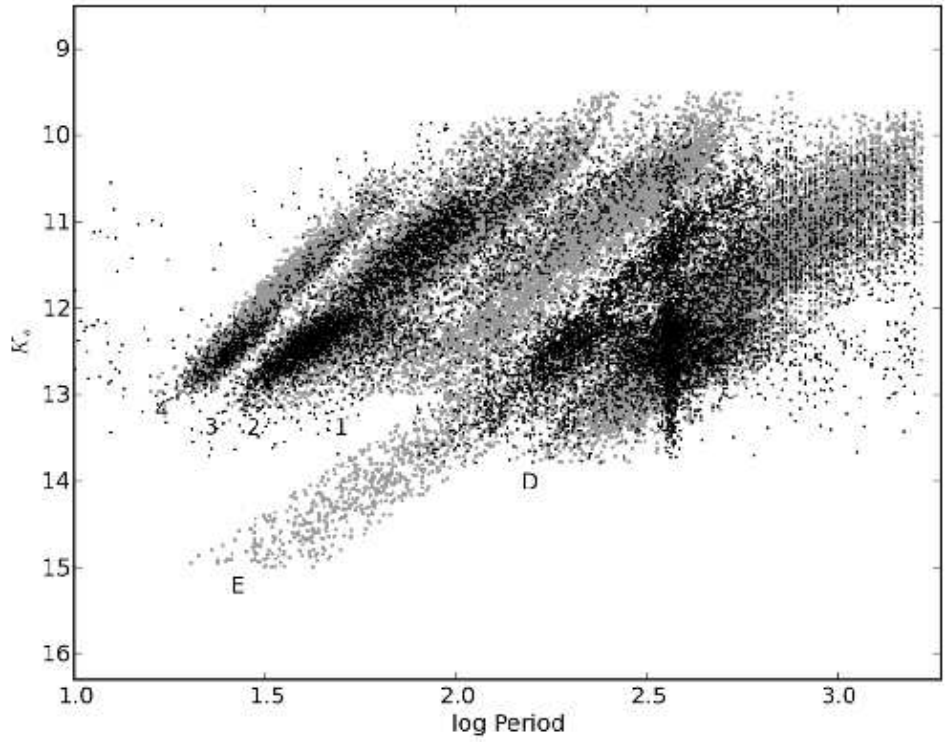}
\caption[Fourier Period-luminosity diagram showing secondary Fourier periods ($P_1$) for stars with their primary Fourier period ($P_0$) on Sequence D.]{Fourier Period-luminosity diagram showing secondary Fourier periods ($P_1$) for stars with their primary Fourier period ($P_0$) on Sequence D. Light gray points show the entire sample of $P_0$ Fourier period-luminosity relations. }
\label{Ds}
\end{center}
\end{figure}

Stars on Sequence D are well-known to be multi-periodic pulsators; \cite{Wood99} found that Sequence D stars exhibited a shorter period that fell on his Sequence B (which is composed of our Sequences 2 and 3). Figure \ref{Ds} shows the secondary Fourier period ($P_1$) of all of the Sequence D stars overlaid on the normal Fourier period-luminosity diagram. We see that many of these periods do in fact lie on Sequences 2 and 3 as found by \cite{Wood99}, although we also see that Sequence 3 is favored. There are also a substantial number of secondary periods of Sequence D stars on Sequence 4 below the tip of the Red Giant Branch, and some between Sequences 1 and 2. However, many of these secondary periods still fall on Sequence D, or have periods half as long as in Sequence D. These stars with $P_0/P_1 \approx 1$ and $P_0/P_1 \approx 2$ are discussed below in the context of the Petersen diagram.

\begin{figure}
\begin{center}
\includegraphics[width=0.9\textwidth]{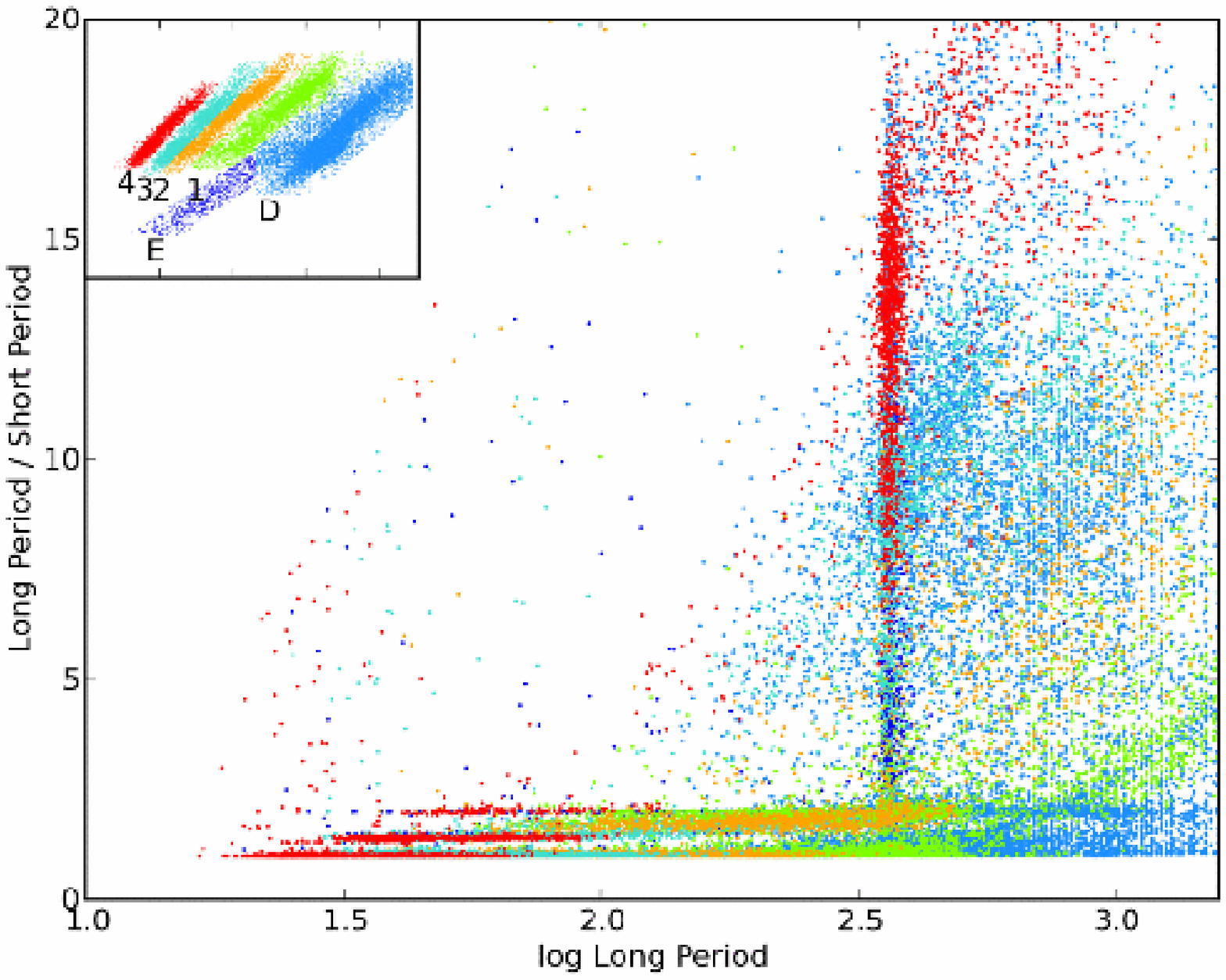}
\caption[Petersen diagram of the LMC]{Petersen diagram of the LMC: for the primary and secondary Fourier periods of a star, the ratio of the longer to the shorter versus the log of the longer period. Stars are color-coded according to the sequence on which their primary Fourier period lies, as shown in the inset. The structure in this diagram is discussed in \S \ref{multiP}. }
\label{Petersen20}
\end{center}
\end{figure}

The Petersen diagram is the plot of the ratio of the two strongest periods versus the longer period in a multi-periodic star, and is presented in Figure \ref{Petersen20}. Multiple periods are found for the great majority of stars in our sample, so stars from all of the Fourier period-luminosity sequences are represented here. The One-Year Artifact is visible in this plot since we only removed stars whose \emph{primary} Fourier periods have low or nonexistent periodicity. Stars with their secondary Fourier periods on the One-Year Artifact lie either in a vertical stripe at $P=\log_{10}(365\ \mathrm{days})$ (when their secondary period is longer than the primary, since it is the longest period that defines placement on the horizontal axis) or along the curve $P_0/P_1 = P_0/(365\ \mathrm{days})$. As observed in \cite{Wood99}, there is a wide locus of points centered around period ratios of 10; these represent stars with their primary or secondary Fourier periods lying on Sequence D. 

\begin{figure}
\begin{center}
\includegraphics[width=0.9\textwidth]{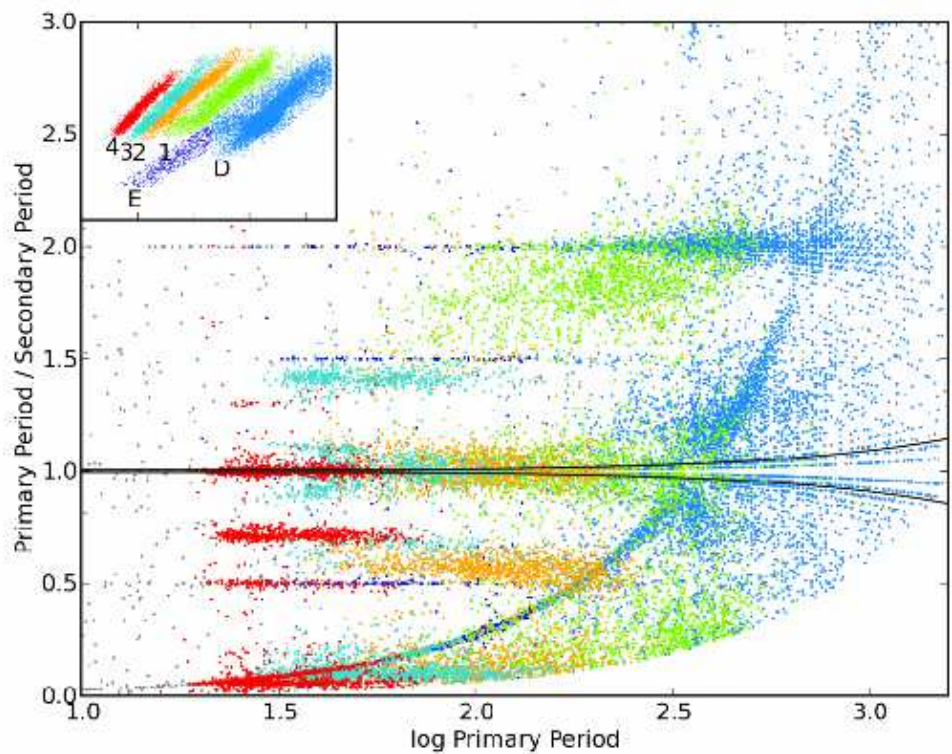}
\caption[$P_0/P_1$ versus $\log_{10}\!P_0$.]{$P_0/P_1$ versus $\log_{10}\!P_0$ showing just the period ratios below 3. Stars are color-coded according to the sequence on which their primary Fourier period ($P_0$) lies, as shown in the inset. Note that $P_0$ is the strongest period, but not necessarily the longest period, as shown by the existence of many stars with period ratios between 0 and 1. Stars with secondary Fourier periods ($P_1$) on the One-Year Artifact lie along the curve $P_0/P_1 = P_0/(365\ \mathrm{days})$. The structure in this diagram is discussed in \S \ref{multiP}. }
\label{Petersen2}
\end{center}
\end{figure}

There is a great deal of structure lying at period ratios below three, which we show in a variant of the Petersen diagram: Figure \ref{Petersen2} plots the $P_0/P_1$ ratio for our stars with respect to the $\log_{10}\!P_0$ Fourier period. Note that $P_0$ is the strongest period, but not necessarily the longest period, as shown by the existence of many stars with period ratios of less than one. In this plot, stars with $P_1$ on the One-Year Artifact lie only along the curve $P_0/P_1 = P_0/(365\ \mathrm{days})$. Aside from stars with periods on the One-Year Artifact, there are at least 10 groups of stars visible with period ratios between zero and two. 

The groups of stars with the smallest period ratios, and therefore the longest secondary periods ($P_1$) relative to their primary period ($P_0$), have their primary periods lying on one of the numbered sequences and their secondary periods lying on Sequence D. 

Many stars show period ratios of approximately one. These closely spaced frequencies are often taken to indicate non-radial pulsation among the least luminous stars from all of the numbered sequences; they have been seen before in both Galactic field stars \citep{2000A&AS..145..283K} and the Magellanic Clouds \citep{2004AcA....54..129S}. The solid lines indicate the region where the period ratio is exactly one within our estimated errors; arcs that are visible at the long-period extreme of this region are due to the limits of our frequency resolution. 

These frequencies can be seen to produce beats, as seen in the light curve of a Sequence 4 star in Figure \ref{LightCurves}, middle panel. It is also possible that some of these closely spaced frequencies may be due to a single, slowing varying period. \cite{2005AJ....130..776T} found that approximately one-tenth of their sample of 547 AAVSO Mira light curves showed ``meandering'' periods that did not simply increase or decrease in length. An example Mira variable from Sequence 1 that has closely spaced frequencies in its CLEANest spectrum, and a changing period, is shown in Figure \ref{LightCurves}, bottom panel. It is overlaid with a sine curve corresponding to the primary Fourier period that CLEANest found: 168.55 days. If the light curve of this star is split in half at the 1760 day mark, we find that the Fourier period of the first half is 166.62 days while the second half has a Fourier period of 163.86 days.

Apart from the period ratios above, stars in Sequences 1 and 2 and stars in Sequences 3 and 4 tend to show secondary periods that correspond to the other Fourier period-luminosity sequence in each group. The Sequence 2 stars at a period ratio of 0.6 have their secondary periods on Sequence 1, while the Sequence 1 stars at ratios between 1.6 and 2.2 have their secondary periods on Sequence 2. Likewise, stars on Sequences 3 and 4 show groups at period ratios of 1.4 and 0.7. We do see stars whose periods cross over between these two groups, for example, the Sequence 4 stars at a period ratio of 0.5, which indicate a secondary period on Sequence 2, and a few Sequence 3 stars at a period ratio of 0.65, indicating a secondary period on Sequence 1. There are even fewer stars that cross from Sequences 1 or 2 to the shorter period sequences, but some Sequence 2 stars are visible at period ratios around two, which represent secondary periods on Sequence 4, and a scattering of Sequence 1 stars lie at period ratios greater than three, which represent secondary periods on Sequence 3. The fact that the period ratios of stars on each pair of Sequences, 3 and 4, and 1 and 2, are reciprocals means that on a Petersen diagram these groups would substantially overlap, indicating that similar physical mechanisms are occurring in each case.

\begin{figure}
\begin{center}
\includegraphics[width=0.9\textwidth]{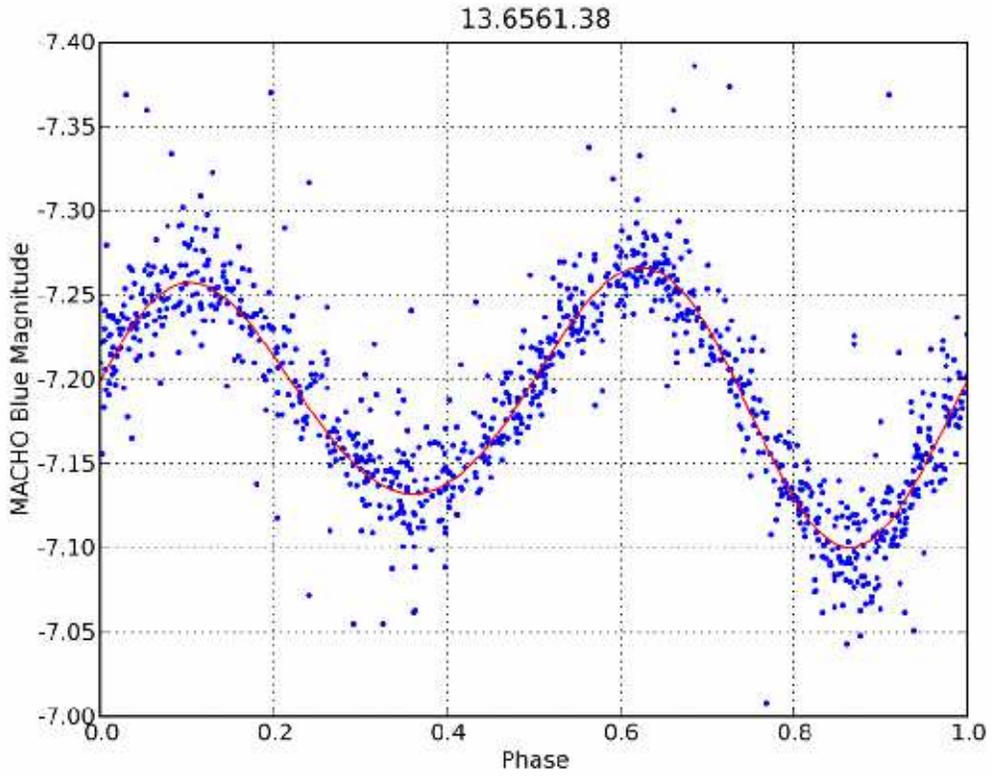}
\caption[Example light curve of a star on Sequence E]{Example light curve of a star on Sequence E with the characteristic alternating minima of a binary system. Note that this light curve is phased at twice $P_0$ period.}
\label{EclipsingBinary}
\end{center}
\end{figure}

Visible at a period ratio of exactly 0.5, 1.5 and 2 are groups of stars from Sequence E. The CLEANest periods found for these contact binaries represent only half of the orbital period of these systems; therefore, these period ratios represent the true orbital period, and the third and fourth harmonics, respectively. These stars often have minima of alternating depth which is characteristic of contact binaries (Figure \ref{EclipsingBinary}). 

The Sequence D stars with $P_0/P_1 \approx 2$ have period ratios of exactly two within our estimated errors, indicating the prevalence of the second harmonic in these light curves. The cause for the closely spaced frequencies ($P_0/P_1 \approx 1$) in Sequence D stars is still unknown at present.

Finally, the sparsely populated fifth sequence \citep{2004AcA....54..129S} is visible in the Sequence 4 stars with ratios of 1.3. 

\subsection{Period Changes}

Period changes are well known for Miras \citep{2005AJ....130..776T} and semi-regular variables \citep{2000A&AS..145..283K}. \cite{2004A&A...425..595G} observed changes from historical periods in OGLE LMC LPVs that indicated stars had moved between Sequences 1, 2, and D\footnote{Actually between Sequences B, C, and D. We take their Sequence B as Sequence 2, but not Sequence 3, because these stars lie on the long period side of Sequence B (Figure 8 from \cite{2004A&A...425..595G}).}. It's likely that some of the stars in our sample underwent a period change during the eight years of the MACHO Project. To explore this possibility, we have split each of our light curves into two halves of equal length and run our analysis on each half independently. The split light curves are almost four years long, which \cite{2005MNRAS.359L..42L} found was a sufficient timespan to resolve the familiar Fourier period-luminosity sequences. This is the case for our split light curves as well, but instead of finding changing periods, this technique appears to find stars with multiple periods of almost equal power.

\begin{figure}
\begin{center}
\includegraphics[width=0.9\textwidth]{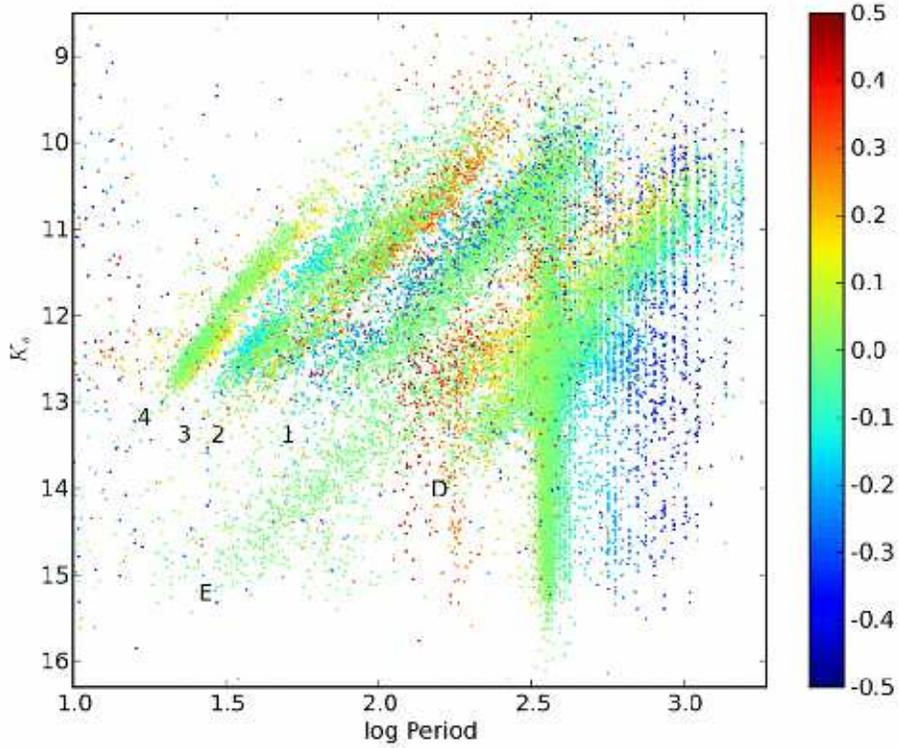}
\caption[Fourier period-luminosity diagram showing period changes.]{Fourier period-luminosity diagram showing period changes of less than 0.5 dex, color-coded by the $\log10$ change in period between the first half of the light curve and the second half. Stars are plotted at the period found in the first half of the light curve. The majority of the period changes of stars in Sequences 4, 3, and 2 are to Sequence D (see Table \ref{sequences}); these larger period changes (which are typically $> 0.5$ dex) are not plotted so that the underlying pattern of smaller period changes is visible. These smaller changes show switching between Sequences 3 \& 4, and Sequences 1 \& 2, but only for stars on the inside edges of these pairs.}
\label{PeriodChanges}
\end{center}
\end{figure}

Fifty percent of this sample shows changes in $\log10$ period of greater than 0.1 dex, the approximate width of the sequences. We have tabulated these movements in Table \ref{sequences} for all cases where more than 10 percent of the stars in one sequence move to another. Many move to Sequence D, and we also see stars switching sequences between the pairs 3 and 4, and 1 and 2. These movements between the pairs of numbered sequences only happen for stars in the inside edge of each pair of sequences, as shown in Figure \ref{PeriodChanges}. 

Considering the multi-periodic nature of these stars, and the much smaller observed rate of period changes in Miras from \cite{2005AJ....130..776T}, we do not believe these data indicate a flurry of period changes. Instead, we observe that the changes we see correspond very well to the multiple periods explored in \S \ref{multiP}. Examination of some of the light curves of these stars shows that both periods are typically present in the frequency spectrum of both halves of the light curve, but the relative amplitudes of these periods change. For stars moving between a short period sequence (such as 3 or 4) and Sequence D we see that the power in the high frequency component becomes distributed among multiple closely-spaced frequencies, which no longer have greater power (individually) than the stable low-frequency component. It's unclear if this is an effect of the different time-sampling in each half of the light curve, or if it is evidence of some change in the star itself.

Although this analysis may not find period changes directly, the equivalence of these periods can be understood as a result of longer term period changes. \cite{1986MNRAS.219..525W} and \cite{agbstars2} argued that the period-luminosity sequences could be understood as the result of different pulsation modes that are excited in turn as the star evolves. The stars that are close to the point where the dominant modes switch (and therefore the sequence switches as well) would be expected to have multiple periods with similar amplitudes

\section{Discussion}
\label{discussion}

The analysis of the frequency spectra of variable stars has been used with great success for characterizing light curve morphology, identifying binary stars, and constraining observed pulsation modes in studies of Cepheids \citep{1981ApJ...248..291S}, Beat Cepheids \citep{1995AJ....109.1653A}, Type II Cepheids and RV Tauri stars \citep{1998AJ....115.1921A}, RR Lyrae \citep{2000ApJ...542..257A}, and LPVs (see \S \ref{intro}). Here we discuss how these techniques contribute to the understanding of the origin of Sequence D and the relationship between the different LPV stages and stellar evolution.

\subsection{Sequence D}

One-third of the stars in our sample exhibit the ``long secondary period'' phenomenon. Although the exact mechanism for Sequence D is still unknown, there are many reasons to think that it is correlated with binary systems. In this paper, the evidence includes the similarity between the luminosity functions of Sequence D and our entire sample (Figure \ref{LumFunctions}, right panel), and the similar amplitudes of Sequence D's average infrared light curves across the 2MASS bands. Other evidence includes the smooth connection between Sequence D and Sequence E---when E is plotted at its orbital period \citep{2004AcA....54..347S}, the presence of ellipsoidal light curves \citep{2007ApJ...660.1486S}, and radial velocity studies of these stars \citep{2006MmSAI..77..537A}. However, there are also several significant features of Sequence D that are unique, or show differences from Sequence E.  Sequence D does not suffer from period halving like Sequence E does, and Sequence E stars often have the third and fourth harmonics present in their frequency spectra, while Sequence D stars show the second harmonic instead (\S \ref{multiP}). \cite{2006ApJ...650L..55D} found that in an amplitude-luminosity plot, stars in Sequence D follow a different pattern than in Sequence E. We see that Sequence E appears to be a continuation of only the small-amplitude ($<0.2$ Blue peak-to-peak) group of stars in Sequence D, those stars that roughly correspond to the OSARGs of \cite{2004AcA....54..129S}. This is consistent with the comparison of the average light curve amplitudes in Blue and $K_s$, where only the lower amplitude Sequence D stars were a good match to Sequence E. Finally, Sequence D's infrared light curves \emph{lead} the optical light curves by 10--15 percent, which is a feature unique to this sequence. These facts do not necessarily preclude a binary star mechanism for Sequence D, but they are useful constraints for proposed mechanisms. We note, as an additional constraint, that stars associated with Sequence 1 are far less likely to have a period lying on Sequence D, and that Sequence D is not observed in the LMC below  $K_s \approx 13.7$.

The model proposed by \cite{2007ApJ...660.1486S}, based on the original model proposed by \cite{Wood99}, is that of a binary system in which the mass lost from the red giant is concentrated near the companion, and regularly obscures the red giant. The wide range of observed light curve amplitudes for Sequence D stars, from 0.1 up to five magnitudes in the MACHO Blue filter \citep{fraser05},  can be readily explained by the projection effects of different inclination angles in a binary system. 

If Sequences E and D are truly composed of binary systems, then the population of binary stars in our sample includes stars with either their primary or secondary Fourier period lying within the boundaries of these sequences (e.g. the top panel of Figure \ref{LightCurves} shows a star whose secondary Fourier period lies on Sequence D). After removing periods identified with the One-Year Artifact, we find that 48 percent of the stars in our sample have variability associated with Sequences E or D (see Table \ref{sequences}). For comparison, \cite{1991A&A...248..485D} found a binary fraction of approximately 40 percent for nearby solar-type stars, and \cite{1997AJ....113.2246R} found a fraction of approximately 35 percent for low-mass stars, a trend in mass which is discussed in \cite{2006ApJ...640L..63L}. It is reasonable to assume that we cannot see pole-on binaries, and there is no reason to think that all binaries have periods shorter than four years, both of which imply that 48 percent underestimates the total percentage of binaries seen in the LMC. This is a serious problem for any explanation of Sequence D that relies solely on binary systems. However, it is very likely that a subset of these stars do show variability due to binarity, perhaps stars in one of the populations that can be separated by color or amplitude.

%Assuming that binary systems are oriented randomly, the range 
% the vector normal to the orbital plan could point anywhere

\subsection{Comparison to Evolutionary Models}

\begin{figure}
\begin{center}
\includegraphics[width=0.9\textwidth]{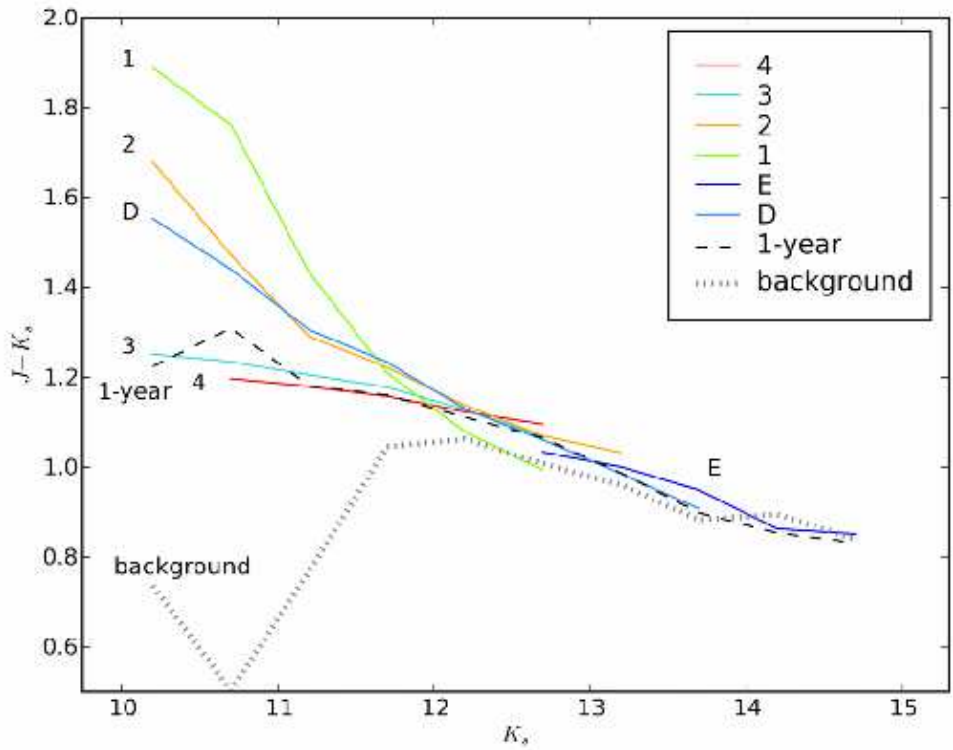}
\caption[Average $J-K_s$ color for each sequence.]{Average $J-K_s$ color for each sequence, as well as stars in the One-Year Artifact and stars in the ``background'' of the Fourier period-luminosity diagram. $J-K_s$ colors are calculated for 0.5 magnitudes bins and shown for bins with more than 40 stars.}
\label{JK}
\end{center}
\end{figure}

We can begin to characterize the evolution of LPVs by comparison in color-magnitude space to models. \cite{2003A&A...403..225M} produced a population synthesis model of the $K_s$ vs. $J-K_s$ color-magnitude diagram (CMD) of the LMC using the RGB and Early AGB (E-AGB) evolutionary models of \cite{2000A&AS..141..371G}, and a preliminary version of their thermally pulsing AGB star models (published later in \cite{2007A&A...469..239M}). Their Figure 12 illustrates the luminosities and colors corresponding to major phases in an LMC star's evolution from RGB, through the Early AGB, and finally to the thermally pulsing AGB (including a transformation to carbon-dominated atmospheres for some stars). For comparison, our Figure \ref{JK} shows the average $J-K_s$ color binned in magnitude for each of the Sequences 1--4, E, D, as well as the One-Year Artifact and the background population of stars. The color of the ``background'' stars match Galactic disk turn-off stars and LMC intermediate-mass stars on the Early AGB in the synthetic CMD of \cite{2003A&A...403..225M}.

As shown in Figure 12, the $J-K_s$ color is more effective at distinguishing evolutionary stages in the AGB than in the RGB. However, with the additional information from the luminosity functions, we can also estimate the importance of RGB stars to each sequence. A distinct peak in the luminosity function at the tip of the RGB  ($K_s=12.3$, \cite{2000ApJ...542..804N}) is widely taken to indicate that the majority of the stars dimmer than this point are themselves on the RGB \citep{2002MNRAS.337L..31I,  2003MNRAS.343L..79K, 2004MNRAS.347L..83K, 2004MNRAS.347..720I, fraser05}.

The stars in our sample that show very weak or nonexistent periodicity (the 24 percent identified with the One-Year Artifact) are predominantly RGB stars. The left two panels of Figure \ref{LumFunctions} show the luminosity functions of stars in the One-Year Artifact with Sequences 3 and 4. Below the tip of the RGB the population of One-Year Artifact stars dominates, suggesting that very weak or non-periodic variability is common among RGB stars. Above the tip of the RGB there is a much closer correspondence between the One-Year Artifact and Sequences 3 and 4. Thus it appears that stars at the dimmest luminosities in our sample vary aperiodically while on the RGB, but most begin to show periodic behavior when they brighten to $K_s \approx 13$.

After passing off of the tip of the RGB, stars may pulsate with shorter periods and lower luminosity as RR Lyrae on the horizontal branch. They become LPVs again as they ascend the Early AGB (i.e. prior to the first thermal pulse or helium shell flash). \cite{2004AcA....54..129S} used the slight offset in period between RGB and AGB stars to show that AGB stars pulsate alongside RGB stars below the tip of the RGB.

Evolution proceeds to brighter luminosities until the onset of thermal pulses, which begin at $K_s \approx 12$ on the synthetic CMD from \cite{2003A&A...403..225M}. Stars primarily populate Sequences 1 and 2 above this luminosity. \cite{1996A&A...311..509W} predict that thermal pulses will create large modulations in luminosity and the pulsation period (and mode) in AGB stars with timescales of thousands of years. The models of \cite{2007A&A...469..239M} substantially agree with these predictions, and show large changes in pulsation period due to mode switching as a direct result of a thermal pulse. Period changes in LPVs are well known \citep{2005AJ....130..776T} but only a small percentage of stars at any one time should be undergoing a thermal pulse due to the short timescales of thermal pulses relative to the long inter-pulse period. The observed period changes of LPVs are not well explained by the effects of thermal pulses alone.

The middle two panels of Figure \ref{LumFunctions} compare the luminosity functions of the numbered sequences. The relative importance of the two giant branches shifts from the RGB to the AGB as we move to Sequences 1 and 2. Additionally, the peak number of stars above the tip of the RGB in each sequence is found at higher luminosity from Sequence 4 to Sequence 1. The OSARG versus SRV/Mira distinction of \cite{2004AcA....54..129S}, by virtue of its definition, roughly corresponds to a division between two pairs of sequences: Sequences 3 and 4, and Sequences 1 and 2. This division is also clearly seen in the observed period ratios (Figure \ref{Petersen2}). Considering the synthetic CMD from \cite{2003A&A...403..225M}, OSARGS are closely related to RGB and E-AGB stars, while the SRV/Mira stars are more closely related to thermally pulsing AGB stars.

At $K_s \approx 11$ the synthetic CMD predicts the formation of carbon stars, and the typical $J-K_s$ colors of each sequence diverge (Figure \ref{JK}). Only Sequences 1, 2, and D redden to the expected $J-K_s$ color of the carbon star tail of \cite{2003A&A...403..225M}. Their models also show that these stars do not evolve in brightness after this stage, so the observed range of carbon star luminosities in our sample may be interpreted as a range of stellar masses. \cite{2007A&A...469..239M} show that only stars between 1.2 and 2.5 $M_\odot$ undergo a dredge-up that can bring the results of nuclear burning to the atmosphere without quickly destroying it through hot bottom burning. Since Miras exist on Sequence 1 at luminosities both above and below $K_s=11$ \citep{fraser05}, not all large amplitude pulsators are carbon stars. Also, not all carbon stars have such red $J-K_s$ colors: \cite{2004A&A...425..595G} found carbon stars on the shorter period sequences---the popular color-cut of $J-K_s \ge 1.4$ appears more effective at segregating M stars, which are rarely this red.

Using $J-K_s>1.4$ to select areas that include only carbon stars, we see that many occupy the highest luminosities of both Sequences 1 and 2, as also seen in \cite{2007arXiv0710.0953L}. Approximately 40 percent of the carbon stars lie on Sequence 2, similar to the prediction of \cite{2003A&A...403..225M}, who fit the observed $K_s$ vs. $J-K_s$ CMD by assuming a 50 percent mix of fundamental and first-overtone pulsation among the carbon stars in their model. Figure \ref{JK} shows that the stars on Sequence 1 evolve to redder $J-K_s$ colors than on Sequence 2, presumably due to increased mass loss driven by fundamental-mode pulsation \citep{2007A&A...469..239M}. Some stars on the long-period extreme of Sequence 1 are underluminous for their color, which may be self-extinction due to dusty outflows.

Beyond this point, stars begin their rapid post-AGB evolution, and they quickly move out of the color-magnitude space of our sample.

\subsection{Brief Summary of LPV Evolution}

LMC stars on the RGB and AGB are characterized by the presence of multiple long periods that show increasing length and amplitude as these stars evolve. Apart from the presence of the long secondary period phenomenon, which appears in approximately half of the stars in our sample, stars initially vary non-periodically, and only later begin to pulsate with periods of 20--120 days on Sequences 3 and 4. The Petersen diagram and its variants show that stars with their primary period on either one of these sequences often have their secondary period on the other sequence. Furthermore, the period changes between the first half and last half of the MACHO light curves suggest that the amplitudes of these pairs of periods are very similar for stars on the inside edges of these two sequences. Similar results are obtained for stars which have evolved to the luminosity at which thermal pulses begin on the AGB ($K_s \approx 12$); these stars are usually found on Sequences 1 and 2 with periods of 45--500 days. This supports the arguments of  \cite{1986MNRAS.219..525W} and \cite{agbstars2}, that a star excites different pulsation modes in turn as it evolves. At the points where the dominant modes switch, we observe pulsation in multiple periods with near equal strengths. The increase in the luminosity of the maximum of the luminosity function above the tip of the RGB also lends support to this argument. The highest luminosity stars on Sequences 1 and 2 have become carbon stars. After this point, the mass-loss rate of these stars increases drastically and they rapidly evolve out of our sample of luminous red stars.

\acknowledgments

This paper utilizes public domain data obtained by the MACHO Project, jointly funded by the U.S. Department of Energy through the University of California, Lawrence Livermore National Laboratory under contract No. W-7405-Eng-48, by the National Science Foundation through the Center for Particle Astrophysics of the University of California under cooperative agreement AST-8809616, and by the Mount Stromlo and Siding Spring Observatory, part of the Australian National University.

This work was performed under the auspices of the U.S. Department of Energy by Lawrence Livermore National Laboratory in part under Contract W-7405-Eng-48 and in part under Contract DE-AC52-07NA27344.

This publication makes use of data products from the Two Micron All Sky Survey, which is a joint project of the University of Massachusetts and the Infrared Processing and Analysis Center/California Institute of Technology, funded by the National Aeronautics and Space Administration and the National Science Foundation.

The Condor Software Program (Condor) was developed by the Condor Team at the Computer Sciences Department of the University of Wisconsin-Madison. All rights, title, and interest in Condor are owned by the Condor Team.

\newpage

\bibliography{../refdb}

\newpage % I've found that, between tables, \clearpage works better than \newpage

\begin{deluxetable}{llrrrr}
\tabletypesize{\scriptsize}
\tablecaption{Fit Parameters of the Fourier Period-Luminosity Sequences.\label{irfits}}
\tablehead{ \colhead{Sequence} & \colhead{Band} & \colhead{$\alpha$} & \colhead{$\beta$} & \colhead{Amplitude\tablenotemark{a}} & \colhead{Optical-IR} \\ & & \colhead{($\mathrm{mag}/\log_{10}\!P$)} & \colhead{(mag)} & \colhead{(peak-to-peak)} & \colhead{Phase Lag} }
\startdata
4 RGB\tablenotemark{b} 
& $W$   & -3.8371 & 17.5432 & 0.08 & \nodata \\
& $J$   & -2.2496 & 16.7511 & \nodata & \nodata \\
& $H$   & -2.5210 & 16.2389 & \nodata & \nodata \\
& $K_s$ & -2.6689 & 16.2291 & \nodata & \nodata \\
4 AGB\tablenotemark{b}
& $W$   & -4.7744 & 18.7935 & 0.07 & \nodata \\
& $J$   & -3.8972 & 18.9384 & \nodata & \nodata \\
& $H$   & -3.9638 & 18.1385 & \nodata & \nodata \\
& $K_s$ & -4.1457 & 18.1713 & \nodata & \nodata \\
3 RGB\tablenotemark{b}
& $W$   & -3.0237 & 16.9699 & 0.08 & \nodata \\
& $J$   & -2.0084 & 16.7729 & \nodata & \nodata \\
& $H$   & -2.3948 & 16.4997 & \nodata & \nodata \\
& $K_s$ & -2.4976 & 16.4552 & \nodata & \nodata \\
3 AGB\tablenotemark{b}
& $W$   & -5.0891 & 20.2000 & 0.17 & \nodata \\
& $J$   & -3.9127 & 19.7273 & \nodata & \nodata \\
& $H$   & -3.9762 & 18.9294 & \nodata & \nodata \\
& $K_s$ & -4.2055 & 19.0728 & \nodata & \nodata \\
2 RGB\tablenotemark{b}
& $W$   & -2.5347 & 16.4754 & 0.17 & \nodata \\
& $J$   & -1.9506 & 16.9278 & \nodata & \nodata \\
& $H$   & -2.2132 & 16.5054 & \nodata & \nodata \\
& $K_s$ & -2.2066 & 16.2846 & \nodata & \nodata \\
2 AGB\tablenotemark{b}
& $W$   & -4.8399 & 20.2378 & 0.41 & \nodata \\
& $J$   & -2.6364 & 17.8648 & 0.07 & 14\% \\
& $H$   & -3.0089 & 17.6513 & 0.06 & 12\% \\
& $K_s$ & -3.6511 & 18.5875 & 0.04 & 14\% \\
1 Oxygen-stars\tablenotemark{c}\\
\hspace{1em}$\mathrm{amp} \le 0.4$ 
& $W$   & -3.9231 & 19.5621 & 0.22 & \nodata \\
& $J$   & -2.9129 & 19.0857 & 0.09 & 24\% \\
& $H$   & -3.0443 & 18.4948 & 0.07 & 23\% \\
& $K_s$ & -3.2104 & 18.5805 & 0.06 & 22\% \\
\hspace{1em}$0.4 < \mathrm{amp} < 1$
& $W$   & -5.7398 & 23.5574 & 0.64 & \nodata \\
& $J$   & -3.3339 & 20.1339 & 0.04 & 3\% \\
& $H$   & -3.4560 & 19.5341 & 0.06 & -1\% \\
& $K_s$ & -3.6736 & 19.7285 & 0.07 & -2\% \\
\hspace{1em}$\mathrm{amp} \ge 1$
& $W$   & -5.1763 & 22.5559 & 1.96 & \nodata \\
& $J$   & -3.2996 & 20.1896 & 0.31 & 11\% \\
& $H$   & -3.4980 & 19.7971 & 0.31 & 13\% \\
& $K_s$ & -3.8292 & 20.2298 & 0.26 & 15\% \\
1 Carbon-stars\tablenotemark{c}\\
\hspace{1em}$\mathrm{amp} \le 0.4$ 
& $W$   & -3.3509 & 18.1327 & 0.28 & \nodata \\
& $J$   & -1.9225 & 17.1136 & 0.12 & 3\% \\
& $H$   & -2.1524 & 16.5671 & 0.09 & 2\% \\
& $K_s$ & -2.5586 & 16.9922 & 0.08 & 16\% \\
\hspace{1em}$0.4 < \mathrm{amp} < 1$
& $W$   & -4.3239 & 20.4330 & 0.62 & \nodata \\
& $J$   & -2.1333 & 17.7452 & 0.18 & 0\% \\
& $H$   & -2.5860 & 17.7218 & 0.13 & 1\% \\
& $K_s$ & -3.1615 & 18.5259 & 0.08 & -1\% \\
\hspace{1em}$\mathrm{amp} \ge 1$
& $W$   & -3.9039 & 19.7674 & 1.80 & \nodata \\
& $J$   & -0.4514 & 13.9507 & 0.58 & 8\% \\
& $H$   & -1.5353 & 15.4351 & 0.49 & 12\% \\
& $K_s$ & -2.8910 & 17.9953 & 0.36 & 13\% \\
E
& $W$   & -2.9765 & 19.1364 & 0.13 & \nodata \\
& $J$   & -2.4438 & 19.1903 & 0.05 & 11\% \\
& $H$   & -2.7049 & 18.9063 & 0.06 & 6\% \\
& $K_s$ & -2.8198 & 18.9370 & 0.05 & 4\% \\
D $\mathrm{amp} \le 0.2$
& $W$   & -2.9953 & 19.7903 & 0.09 & \nodata \\
& $J$   & -2.4171 & 19.6843 & 0.05 & -11\% \\
& $H$   & -2.5615 & 19.1820 & 0.05 & -11\% \\
& $K_s$ & -2.7983 & 19.5482 & 0.05 & -12\% \\
D $\mathrm{amp} > 0.2$
& $W$   & -3.3347 & 20.1750 & 0.48 & \nodata \\
& $J$   & -1.8981 & 18.0679 & 0.08 & -7\% \\
& $H$   & -2.1999 & 17.9674 & 0.08 & -13\% \\
& $K_s$ & -2.5943 & 18.7166 & 0.06 & -16\% \\
\enddata
\tablenotetext{a}{Amplitudes listed after $W$ are the CLEANest amplitudes for the MACHO Blue light curve.}
\tablenotetext{b}{The stars above and below the tip of the Red Giant Branch ($K_s=12.3$) were fit separately for Sequences 2, 3, and 4.}
\tablenotetext{c}{The oxygen and carbon stars in Sequence 1, separated using a color cut at $J-K_s=1.4$, were fit individually.}
\end{deluxetable}

\begin{deluxetable}{lrrccl}
\tabletypesize{\scriptsize}
\tablecolumns{6} % necessary for \cutinhead to work
\rotate          % landscape!
\tablewidth{0pt} % use the actual width of the table instead of the horizontal page width
\tablecaption{Characteristics of the Fourier Period-Luminosity Sequences.\label{sequences}}
\tablehead{ 
\colhead{Sequence} & \multicolumn{2}{c}{Number} & \colhead{Optical/$K_s$} & \colhead{Optical/$K_s$} & \colhead{Period}\\ 
& \multicolumn{2}{c}{of Stars\tablenotemark{a}} & \colhead{Amplitude Ratio} & \colhead{Phase Lag} & \colhead{Changes\tablenotemark{b}} }
\startdata
 4                  &  3,943  &  8\% & \nodata & \nodata & 40\% do not move, 36\% to D, 14\% to 3\\
 3                  &  4,228  &  9\% & \nodata & \nodata & 46\% to D, 30\% do not move, 13\% to 4\\
 2                  &  3,917  &  8\% &      11 &    14\% & 40\% do not move, 26\% to D, 22\% to 1\\
 1                  &  5,485  & 12\% &    4--9 & -2--22\%& 60\% do not move, 16\% to 2, 15\% to D\\
 E                  &    888  &  2\% &       2 &     4\% & 77\% do not move\\
 D                  & 14,438  & 31\% &    2, 8 & -12, -16\%& 70\% do not move\\
One-Year Artifact   & 11,215  & 24\% & \nodata & \nodata \\
unclassified LPVs   &  2,658  &  6\% & \nodata & \nodata\\
\cutinhead{Number of stars with either $P_0$ or $P_1$ on Sequence E, Sequence D, or either Sequence E or D.}
 E         &  1,624  &  4\% & &\\
 D         & 20,805  & 44\% &&\\
E or D     & 22,261  & 48\% &&\\
\enddata
\tablenotetext{a}{The percentages are with respect to all those stars with good Red and Blue CLEANest results and a 2MASS $K_s$ magnitude: 46,831 stars in total.}
 \tablenotetext{b}{From the first half of the light curve to the second half; the listed movements are just those which involve at least 10 percent of the stars in each sequence.}

\end{deluxetable}

\end{document}